\documentclass[journal]{IEEEtran}
 
 \usepackage[T1]{fontenc}
  
 \usepackage{amsmath,amssymb,amsfonts,amsthm,bm,mathtools,dsfont,nccmath}
 \usepackage{graphicx,epsfig,epstopdf}
 \usepackage{caption,subcaption}
 \usepackage{float,stfloats}
 
 \usepackage{algorithm,algpseudocode,algcompatible}
 \algnewcommand\algorithmicreturn{\textbf{return}}
 \algnewcommand\RETURN{\State \algorithmicreturn}
 
 \usepackage{tabularx,enumerate,xspace,enumitem}
 
 \usepackage[sort,compress]{cite}  
 \usepackage[hidelinks]{hyperref}  
 
 \theoremstyle{plain}

 \usepackage{balance,flushend}
 
 \usepackage{xcolor}

 \usepackage[framemethod=tikz]{mdframed}
 \definecolor{mycolor}{rgb}{0.122, 0.435, 0.698}
 \newmdenv[innerlinewidth=0.5pt, roundcorner=4pt, linecolor=mycolor,
 innerleftmargin=6pt, innerrightmargin=6pt,
 innertopmargin=6pt, innerbottommargin=6pt]{mybox}
 
 \newcommand{\bmk}{\mathbf{k}}  
   
   \newcommand{\bmV}{\mathbf{V}}

   \newcommand{\bmW}{\mathbf{W}}
 \newcommand{\bmx}{\mathbf{x}}  
   
 \newcommand{\bmr}{\mathbf{r}}  
 
 \usepackage{ulem}
 \usepackage[user]{zref}
 
 \newcounter{revc}
 \makeatletter
 \zref@newprop{revcontent}{} \zref@addprop{main}{revcontent}
 \zref@newprop{revsec}{} \zref@addprop{main}{revsec}
 \zref@newprop{revpage}{} \zref@addprop{main}{revpage}
 
 \newcommand{\revi}[2]{%
 	\zref@setcurrent{revsec}{\thesection}%
 	\zref@setcurrent{revpage}{\thepage}%
 	\zref@setcurrent{revcontent}{#2}%
 	\refstepcounter{revc}%
 	\label{#1}%
 	\zlabel{#1}%
 	\textcolor{blue}{#2}%
 }
 
 \newcommand{\revinu}[2]{%
 	\zref@setcurrent{revsec}{\thesection}%
 	\zref@setcurrent{revcontent}{#2}%
 	\refstepcounter{revc}%
 	\zlabel{#1}%
 	\label{#1}
 	#2 }
 
 \newcommand{\revr}[2]{%
 	\zref@setcurrent{revsec}{\thesection}%
 	\zref@setcurrent{revcontent}{#2}%
 	\refstepcounter{revc}%
 	\zlabel{#1}%
 	\label{#1} \sout{#2}} 
 \makeatother

 \expandafter\def\expandafter\quote\expandafter{\quote\onehalfspacing\fontsize{12}{14}\selectfont}
 
 \def\BibTeX{{\rm B\kern-.05em{\sc i\kern-.025em b}\kern-.08em
 		T\kern-.1667em\lower.7ex\hbox{E}\kern-.125emX}}

\begin{document}

\title{Joint Holographic Beamforming and User Scheduling with Individual QoS Constraints}  
\author{Chandan~Kumar~Sheemar,~\IEEEmembership{Member,~IEEE}, Christo Kurisummoottil Thomas,~\IEEEmembership{Member,~IEEE},\\
George C. Alexandropoulos,~\IEEEmembership{Senior Member,~IEEE},  Jorge Querol,~\IEEEmembership{Member,~IEEE},\\ Symeon Chatzinotas,~\IEEEmembership{Fellow Member,~IEEE} and Walid Saad,~\IEEEmembership{Fellow Member,~IEEE}
\thanks{Chandan Kumar Sheemar, Jorge Querol and Symeon Chatzinotas are with the SnT department at the University of Luxembourg (email:\{chandankumar.sheemar,jorge.querol,symeon.chatzinotas\}@uni.lu). George C. Alexandropoulos is with the Department of Informatics and Telecommunications, National and Kapodistrian University of Athens, 15784 Athens, Greece (email: alexandg@di.uoa.gr). Christo Kurisummoottil Thomas and Walid Saad are with Virginia Tech \{christokt,walids\}@vt.edu) } 
} 
 \maketitle

\begin{abstract}
 Reconfigurable holographic surfaces (RHS) have emerged as a transformative material technology, enabling dynamic control of electromagnetic waves to generate versatile holographic beam patterns. This paper addresses the problem of joint hybrid holographic beamforming and user scheduling under per-user minimum quality-of-service (QoS) constraints, a critical challenge in resource-constrained networks. However, such a problem results in mixed-integer non-convex optimization, making it difficult to identify feasible solutions efficiently. To overcome this challenge, we propose a novel iterative optimization framework that jointly solves the problem to maximize the RHS-assisted network sum-rate, efficiently managing holographic beamforming patterns, dynamically scheduling users, and ensuring the minimum QoS requirements for each scheduled user. The proposed framework relies on zero-forcing digital beamforming, gradient-ascent-based holographic beamformer optimization, and a greedy user selection principle. Our extensive simulation results validate the effectiveness of the proposed scheme, demonstrating their superior performance compared to the benchmark algorithms in terms of sum-rate performance, while meeting the minimum per-user QoS constraints.
\end{abstract}

\IEEEpeerreviewmaketitle
 
\section{Introduction} \label{Intro}

\IEEEPARstart{T}{he} sixth-generation (6G) wireless communication systems mark a transformative milestone, designed to address the escalating demands for unprecedented spectral efficiency, energy efficiency, and operational flexibility \cite{wang2024tutorial,kamruzzaman2022key,ji2021several}. As global connectivity evolves, 6G networks must meet the diverse needs of users, adapt to highly dynamic environments, and handle densely populated network conditions—all while delivering ultra-high data rates, minimal latency, and exceptional reliability \cite{cao2023toward}. Achieving these ambitious goals requires advanced and sustainable technologies capable of intelligently adapting to ever-changing channel conditions and user demands. Traditional phased array antennas (PAAs) \cite{sheemar2022practical,payami2016hybrid}, which have been foundational in earlier wireless systems, face significant limitations in this context. Their reliance on complex and bulky mechanical hardware for beam steering results in high costs and significant energy consumption, hindering scalability and dynamic adaptability \cite{ikram2022road}. 

To address the aforedescribed limitations, reconfigurable holographic surfaces (RHSs) have emerged as a transformative alternative, redefining the role of antenna front-ends in wireless communication systems \cite{huang2020holographic,gong2023holographic,AXN2023,BAL_VTM_2024,iacovelli2024holographic,HMIMO_computing}. By utilizing reconfigurable metamaterials and the holographic principle, RHS-assisted systems eliminate the dependence on bulky mechanical components and power-intensive phase-shifting circuits, offering compact, lightweight, and energy-efficient solutions \cite{black2017holographic}. Such systems can enable superior flexibility in reconfiguration, spatial resolution, and adaptability \cite{zhang2022holographic}. Unlike traditional PAAs, which face scalability challenges due to high energy demands and intricate mechanical designs, RHS-based systems can function as advanced antenna front-ends, leveraging passive or semi-passive elements to dynamically and efficiently control electromagnetic waves~\cite{shlezinger2021dynamic}. This innovative approach not only improves energy efficiency but also simplifies hardware architecture, reducing overall system complexity and production costs by minimizing reliance on expensive radio frequency (RF) components and active circuitry \cite{lin2019hybrid,di2024reconfigurable}. These efficiencies make RHS systems particularly well-suited for 6G systems \cite{gong2023holographic,deng2022hdma,ghermezcheshmeh2023parametric}.

The compact and innovative design of RHS, fabricated using printed circuit board (PCB) technology, functions on the following principle. Firstly, it integrates a feed mechanism that generates an electromagnetic reference wave \cite{huang2020holographic}. This wave propagates across the metasurface, radiating energy through its array of reconfigurable elements. Using engineered metamaterials, the metasurface creates a holographic pattern by exploiting the interference between the reference wave and a synthesized target wave. Each element of the surface modulates this holographic pattern by dynamically adjusting the holographic weights of the radiation patterns to form highly directional and adaptable beams. This technique, referred to as holographic beamforming, replaces the reliance on traditional PAAs \cite{sheemar2022practical,yetis2021joint,rosson2019towards}, thus leads to a more sustainable wireless ecosystem \cite{wu2018hybrid}.

Although RHS-based systems offer significant advantages \cite{di2021reconfigurable}, their research is still in the early stages, and they face a major challenge compared to traditional PAAs-based transceiver design. Namely, the challenges lies in the methodologies for beamforming techniques, which are typically well established in conventional PAAa-based systems with phase-shifting mechanisms \cite{wu2018hybrid}. However, the holographic beamforming design problems are subject to real valued constraints on the holographic weights, while being a function of the complex variables \cite{deng2022reconfigurable}, thus leading to unconventional problem structures and driving the quest for novel optimization strategies.

Despite these challenges, substantial progress has been made in developing innovative solutions that pave the way for the broader adoption of RHS in wireless communication systems. For example, in \cite{deng2021reconfigurable}, the problem of hybrid holographic beamforming for sum-rate maximization in a RHS-based systems is tackled. The study introduces a solution by jointly optimizing the digital beamformer and the analog holographic beamformer. For the latter, the non-convex optimization problem is approached by breaking it down into a series of convex subproblems, which is iteratively solved. In \cite{hu2023holographic}, an alternative solution is proposed for broadcasting only one data stream to all the users, that employs fully analog holographic beamforming to maximize the sum-rate without the digital beamforming. This approach eliminates the need for expensive RF chains which reduces the cost and power consumption. However, such a solution offers less flexibility for data multiplexing and interference management Further extending the understanding of RHS systems, \cite{an2023tutorial} delves into the spatial degrees of freedom (DoF) and ergodic capacity of a point-to-point RHS setup. By analyzing the fundamental capacity limits of RHS, this work provides important theoretical benchmarks that can guide the design of future systems. In addition to these works, \cite{you2022energy} explores the energy-efficiency maximization problem in RHS systems, focusing on the case where only analog holographic beamforming is employed. The study proposes solutions under both perfect and imperfect channel state information (CSI) scenarios, emphasizing the practical importance of robust and efficient system designs in real-world settings. In \cite{shlezinger2019dynamic,shlezinger2021dynamic}, the authors conducted an in-depth analysis of the operational principles associated with dynamic metasurface antennas (DMAs), which represent a metasurface-based implementation framework for HMIMO transceivers. These studies further explored the advantages and capabilities of DMAs in the context of 6G communications, highlighting their potential and identifying key challenges arising from their broad range of prospective applications. The study in \cite{zhu2024electromagnetic} explored electromagnetic (EM) information theory inspired by HMIMO system design, presenting foundational principles of this interdisciplinary framework, including modelling, analytical methods, and practical applications. In \cite{wan2021terahertz}, the potential of HMIMO for terahertz (THz) communications is investigated. 

Further studies exploring the potential of RHS-based holographic systems for near-field communications have demonstrated significant advancements in this area \cite{XYA2024,ji2023extra,gan2022near,gong2024holographic,wei2023tri,GIS2023,GA2024,GA2024a}. In \cite{ji2023extra}, the authors investigate spatially constrained antenna apertures with rectangular symmetry, focusing on enhancing spatial degrees of freedom and channel capacity. Their approach leverages evanescent waves for information transmission in near-field scenarios, utilizing the Fourier plane-wave series expansion to achieve these improvements. In \cite{gan2022near}, the authors study the integration of RHS into millimeter-wave systems for advanced beamforming and localization. Building on the principles of near-field wavefront manipulation, they explore its potential to enhance both communication and positioning accuracy by optimizing beamforming patterns and minimizing localization errors. In \cite{gong2024holographic}, the authors studied a generalized EM-domain near-field channel model and its capacity limits for a point-to-point holographic system with arbitrarily placed surfaces in a line-of-sight (LoS) environment. They proposed two efficient channel models for continuous RHS—one offering high precision with a sophisticated formulation and another providing computational simplicity with minimal accuracy trade-offs—and derived a tight upper bound for the system capacity using a well-constructed analytical framework. In \cite{ji2023extra}, the authors studied the use of triple polarization (TP) for multi-user wireless communication systems with RHS to enhance capacity and exploit diversity without increasing antenna array size. They developed a TP near-field channel model for continuous RHS based on the dyadic Green's function and proposed two precoding schemes to mitigate cross-polarization and inter-user interference. For further contributions on near-field holographic joint communication and sensing, we refer the reader to \cite{GIS2023,GA2024,GA2024a}.

\subsection{Motivation and Main Contributions}
The integration of joint holographic beamforming and user scheduling under Quality of Service (QoS) constraints is a critical challenge for next-generation wireless systems. As the demand for high data rates and reliable connectivity continues to grow, especially in resource-constrained environments, it becomes essential to design systems that can efficiently allocate resources while ensuring that each served user’s QoS requirements are met. In this context, RHS-based systems, with their flexibility in beamforming and ability to dynamically adjust to varying channel conditions, present a promising solution. However, addressing both the need for high spectral efficiency and stringent QoS constraints simultaneously in an RHS-assisted resource contained network is a complex task. This makes the problem of joint holographic beamforming and user scheduling even more critical, as it directly impacts the network’s ability to provide high performance while meeting the service requirements of each scheduled user.

Despite its importance, the problem of joint hybrid holographic beamforming and user selection under QoS constraints remains largely unexplored in the existing literature. While previous studies have focused on individual components of RHS optimization, such as hybrid beamforming for sum-rate maximization \cite{deng2021reconfigurable}, fully analog beamforming \cite{hu2023holographic, you2022energy}, and capacity analysis for point-to-point communication \cite{an2023tutorial, gong2024holographic}, they have not yet addressed the integration of user scheduling under QoS constraints with holographic beamforming. This lack of exploration is a significant gap, as optimizing both aspects together can unlock the full potential of RHS technology in real-world scenarios, where resource limitations and stringent QoS requirements are prevalent. By combining efficient user selection with holographic beamforming, it is possible to improve system throughput, reduce power consumption, and ensure robust performance even in demanding network conditions.


We address this problem by optimizing the performance of an RHS-assisted multi-user downlink communication system for a hybrid holographic transceiver. Specifically, we focus on the joint optimization of hybrid holographic beamforming and user scheduling to maximize the system’s sum rate, while ensuring that each selected user satisfies its minimum quality-of-service (QoS) requirement, along with the real-domain constraints on holographic weights and the total power budget. The formulated problem results in a mixed-integer non-convex, presenting significant challenges for finding a feasible solution.

To address this problem, we propose a novel algorithmic framework comprising three key components: 1) Lower-dimensional digital beamforming optimization, 2) Higher-dimensional holographic analog beamformer optimization, and 3) QoS-aware user scheduling. For the digital beamforming, the Zero-Forcing (ZF) technique is employed to mitigate inter-user interference while adhering to the power constraints. For the optimization of the holographic analog beamformer, we introduce a novel gradient ascent-based algorithm to dynamically adjust the holographic weights to shape the holographic radiation patterns to minimize interference and enhance signal focus. The algorithm relies on the closed-form expressions for the gradients of the holographic weights, which are first derived. Finally, for user scheduling, a novel greedy principle-based scheme is developed to iteratively select users that maximize the system’s overall performance while satisfying the QoS requirements. This scheme prioritizes users based on their channel conditions and potential interference contributions, dynamically optimizing the active user set to achieve a high sum-rate. 

Simulation results are presented to validate the effectiveness of the proposed framework. The results demonstrate that the framework consistently satisfies the minimum QoS requirements for all selected users while achieving significant improvements in the network's sum rate. These findings underscore the potential of the proposed solution for practical deployment in RHS-assisted next-generation wireless communication systems, particularly in scenarios with limited resources and stringent QoS requirements.

\begin{figure*}
    \centering
   \includegraphics[width=16cm,height=6cm]{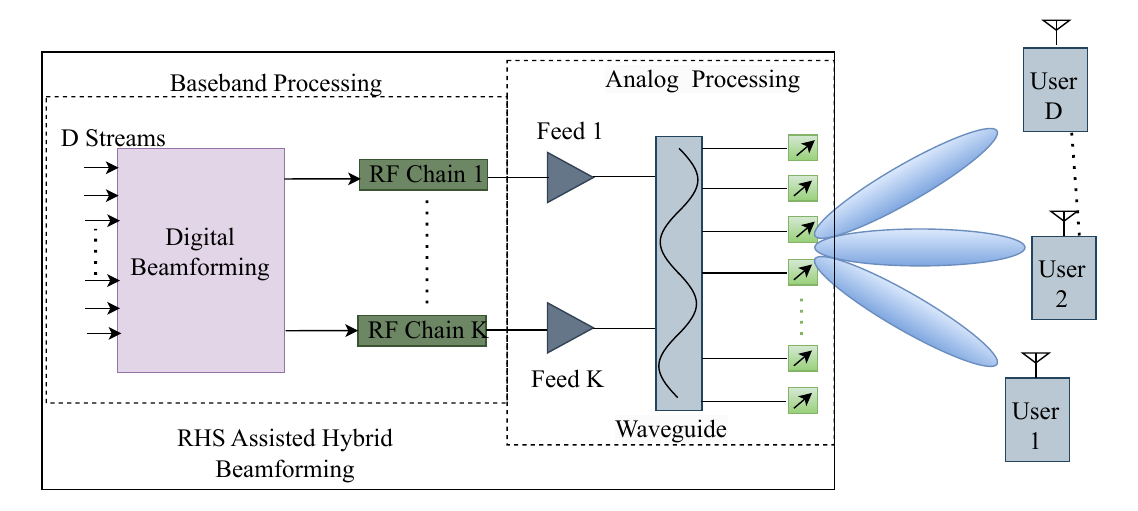}
   \caption{Hybrid RHS-assisted system with digital and analog holographic beamforming.}
    \label{fig1}
\end{figure*}
\emph{Paper Organization:} The remainder of this paper is organized as follows. Section \ref{section_2} presents the system model and formulates the optimization problem for joint hybrid holographic beamforming and user scheduling in an RHS-assisted system with QoS constraints. Section \ref{section_3} presents a novel algorithmic framework to solve the joint problem. Finally, Section \ref{section_4} presents the simulation results, and Section \ref{section_5} concludes the paper.

\emph{Notations:} In this paper, we adopt a consistent set of notations: Scalars are denoted by lowercase or uppercase letters, while vectors and matrices are represented by bold lowercase and bold uppercase letters, respectively. The transpose, Hermitian transpose, and inverse of a matrix $\mathbf{A}$ are denoted by $\mathbf{A}^\mathrm{T}$, $\mathbf{A}^\mathrm{H}$, and $\mathbf{A}^{-1}$, respectively. Sets are indicated by calligraphic letters (e.g., $\mathcal{A}$), and their cardinality is represented by $|\mathcal{A}|$. Finally, $\|\cdot\|$ denotes the indicate the $l_2$-norm.  

\section{System Model} \label{section_2}
 
\subsection{Scenario Description}

We consider a multi-user downlink communication system as shown in Figure \ref{fig1}, where a base station (BS) transmits data streams to $D$ mobile users, indexed by the set $\mathcal{D} = \{1, \dots, D\}$. Each user is assumed to have a minimum QoS requirement denoted as $R_d^{\textrm{min}}, \forall d \in \mathcal{D}$. The BS employs a hybrid beamforming architecture, which consists of two components: a high-dimensional holographic beamformer in the analog domain and a lower-dimensional digital beamformer. The digital beamforming matrix is represented as $ \mathbf{V} \in \mathbb{C}^{R \times D} $, where $R$ denotes the number of RF chains. Each column of $\mathbf{V}$, denoted by $\mathbf{v}_d \in \mathbb{C}^{R \times 1}$, corresponds to the digital beamformer for user $d \in \mathcal{D}$. The number of RF chains in the hybrid RHS architecture is assumed to satisfy $ R \geq D $, ensuring that the system can support the active data streams. Each RF chain feeds its corresponding RHS feed, which transforms the high-frequency electrical signal into a propagating electromagnetic reference wave. Let $K$ denote the number of feeds and we assume the number of RF chains to be equal to the number of feeds. The radiation amplitude of the reference wave at each metamaterial element of the RHS is dynamically controlled using a holographic beamforming matrix $ \bmW \in \mathbb{C}^{M \times K} $, enabling the generation of desired directional beams. We assume a rectangular RHS having $\sqrt{M}$ elements along the x-axis and  $\sqrt{M}$ elements along the y-axis, and the total number of RHS elements are given as $M$. Each element of the holographic matrix $\bmW$ is of the form $\{w_m \cdot e^{-j \bmk_s \cdot \bmr_{m}^k}\}$ \cite{deng2021reconfigurable}, where $e^{-j \bmk_s \cdot \bmr_{m}^k}$ is the phase of the reference wave with $\bmk_s$ denoting the reference wave vector and $\bmr_{m}^k$ denoting the distance from the $k$-th feed to the $m$-th element, and $0 < m \leq M$.

Let $\mathbf{h}_d \in \mathbb{C}^{M \times 1}$ denote the channel response for the user $d$ and let $y_d$ denote its received signal. Based on the aforementioned notations, we can write $y_d$ as follows
\begin{equation}
    y_d = x_d \underbrace{\mathbf{h}_d^H \bmW \mathbf{v}_d s_d}_{\text{Desired Signal}} + \underbrace{\sum_{k \neq d} x_k \mathbf{h}_d^H \bmW \mathbf{v}_k s_k}_{\text{Inter-User Interference}} + \underbrace{n_d}_{\text{Noise}},
\end{equation}
where $ s_d $ is the transmitted symbol for user $ d $, assumed to be i.i.d. with unit variance, i.e., $ \mathbb{E}[s_d s_d^*] = 1 $, and $ n_d \sim \mathcal{CN}(0, \sigma^2) $ is the additive white Gaussian noise (AWGN) at user $ d $. Let $\mathcal{S} = \{0,1\}$ denote the set containing the scheduling status of the users. The scaler $x_i \in \mathcal{S}$ denotes a binary variable which assumes value $1$ if user $i$ is scheduled and $0$ otherwise.

The channel between the base station and user $ d $ is denoted as $\mathbf{h}_d \in \mathbb{C}^{M \times 1} $, which is modelled as a sparse millimeter-wave (mmWave) channel with $ I $ propagation paths. The channel can be expressed as \cite{deng2021reconfigurable}
\begin{equation}
    \mathbf{h}_d = \sqrt{\frac{M}{I}} \sum_{i=1}^{I} \alpha_i^d  \mathbf{a}_t(\theta_i^d, \varphi_i^d),
\end{equation}
where $ \alpha_i^d $ represents the complex gain of the $ i $-th path. The term $ \mathbf{a}_t(\theta_i^d, \varphi_i^d) $ is the transmit array response vector corresponding to the angles of departure (AoDs) $ \theta_i^d $ (elevation) and $ \varphi_i^d $ (azimuth) of the $ i $-th path. We model the response of the RHS as a uniform planner array (UPA) of size $\sqrt{M} \times \sqrt{M}$, where $ \sqrt{M}$ and represent the number of antenna elements along the $ x $-axis and $ y $-axis. The array response vector can be written as
\begin{equation}
    \mathbf{a}_t(\theta^d, \varphi^d) = \frac{1}{M}\mathbf{a}_x(\theta^d, \varphi^d) \otimes \mathbf{a}_y(\theta^d, \varphi^d),
\end{equation}
where $ \otimes $ denotes the Kronecker product, and the components $ \mathbf{a}_x(\theta, \varphi) $ and $ \mathbf{a}_y(\theta, \varphi) $ are the steering vectors for the $ x $- and $ y $-axes, given by:
\begin{equation}
    \mathbf{a}_x(\theta^d, \varphi^d) = 
    \begin{bmatrix}
    1 \\
    e^{j k_f d_x \sin\theta^d \cos\varphi^d} \\
    e^{j 2 k_f d_x \sin\theta^d \cos\varphi^d} \\
    \vdots \\
    e^{j (\sqrt{M}-1) k_f d_x \sin\theta^d \cos\varphi^d}
    \end{bmatrix},
\end{equation}
\begin{equation}
    \mathbf{a}_y(\theta^d, \varphi^d) = 
    \begin{bmatrix}
    1 \\
    e^{j k_f d_y \sin\theta^d \sin\varphi^d} \\
    e^{j 2 k_f d_y \sin\theta^d \sin\varphi^d} \\
    \vdots \\
    e^{j (\sqrt{M}-1) k_f d_y \sin\theta^d \sin\varphi^d}
    \end{bmatrix}.
\end{equation}
where $ k_f$ denotes the wavenumber in free space, and $ d_x $ and $ d_y $ denote the spacing between adjacent elements along the $ x $- and $ y $-axes, respectively.

\subsection{Optimization Problem}

To evaluate the performance of the RHS-assisted multi-user system, we define the Signal-to-Interference-plus-Noise Ratio (SINR) for user $ d $ as a measure of the signal quality at the receiver. It quantifies the ratio of the desired signal power to the interference and noise power, and for user $ d $, it is denoted as $\text{SINR}_d$ and given as
\begin{equation}
    \text{SINR}_d = \frac{\left|x_d \mathbf{h}_d^H \bmW \mathbf{v}_d \right|^2}{\sigma^2 + \sum_{k \neq d} \left| x_k\mathbf{h}_d^H \bmW \mathbf{v}_k \right|^2},
\end{equation}
where the numerator represents the power of the desired signal, and the denominator consists of the noise power $ \sigma^2 $ along with the interference from all other users $ k \neq d $, captured by the second term. The binary variable $ x_i $ indicates the scheduling status of user $ i $, where $ x_i = 1 $ signifies that the user is scheduled, and $ x_i = 0 $ denotes that the user is inactive.

We aim to maximize the system's sum rate by jointly optimizing the digital beamforming matrix $ \mathbf{V} = [\mathbf{v}_1, \dots, \mathbf{v}_D] $, where each $ \mathbf{v}_d $ represents the beamformer for user $ d $, the holographic beamformer matrix $ \bmW $, and the binary user scheduling vector $ \mathbf{x} = [x_1, \dots, x_D] $, with $ x_d $ indicating the scheduling status of user $ d $.  The problem is subject to several constraints: the total sum-power constraint, the real-domain constraint on the elements of the holographic beamformer $ \bmW $ \cite{deng2021reconfigurable}, the binary nature of the user scheduling variables $ x_d $, and the minimum QoS requirement for each selected user. 

The joint optimization problem for an RHS-assisted system under the aforementioned constraints can be formulated as
\begin{subequations} \label{problem_statement}
    \begin{equation}
    \begin{aligned}
          \max_{\mathbf{V}, \bmW,\mathbf{x}} \quad & \sum_{d=1}^D \log_2 \left( 1 + \frac{\left|x_d \mathbf{h}_d^H \bmW \mathbf{v}_d \right|^2}{\sigma^2 + \sum_{k \neq d} \left| x_k\mathbf{h}_d^H \bmW \mathbf{v}_k \right|^2} \right),
    \end{aligned}
    \end{equation}
    \begin{equation} \label{c1}
        \textbf{s.t.} \quad   \text{Trace}(\bmV \bmV^H) \leq P_{\text{max}},
    \end{equation}
    \begin{equation} \label{c2}
        \quad  \quad  x_d \in \mathcal{S}, \quad \forall d \in \mathcal{D},
    \end{equation}
    \begin{equation} \label{c3}
     \hspace{9mm} \log_2\left(1 + \frac{\left|x_d \mathbf{h}_d^H \bmW \mathbf{v}_d \right|^2}{\sigma^2 + \sum_{k \neq d} \left| x_k\mathbf{h}_d^H \bmW \mathbf{v}_k \right|^2}\right) \geq R_d^{\text{min}},   \forall x_d =1,
    \end{equation}
    \begin{equation} \label{c4}
        0 \leq w_{m} \leq 1,  \quad  \forall\; 0 < m \leq M. 
    \end{equation}
\end{subequations} 
 
where $\eqref{c1}-\eqref{c2}$ denote the sum-power and the binary constraint for the scheduling variable, respectively, and $\eqref{c3}-\eqref{c4}$ denote the minimum QoS constraint for the scheduled users and the real-valued constraint on the holographic beamformer.

 The optimization problem is highly complex and results to be mixed integer non-convex due to several key factors. The objective function involves a sum of logarithmic terms based on SINR, which is non-linear and depends on quadratic terms in the beamforming matrix $ \mathbf{V} $. The inclusion of binary variables $x_d$ for user selection introduces combinatorial complexity, while the power constraint $ \text{Trace}(\bmV \bmV^H) \leq P_{\text{max}} $ and minimum rate requirements add further non-convexity. 
 Furthermore, note that problem \eqref{problem_statement} differs significantly from conventional phase-controlled beamforming design as it involves complex-domain optimization subject to unconventional real-domain amplitude constraints, see e.g. \cite{deng2021reconfigurable}. These constraints are further complicated by the superposition of radiation waves emitted from different radiation elements. Traditional algorithms with fully digital beamforming only \cite{he2022joint,huang2022joint} or hybrid beamforming are not well-suited for this scenario, as the analog beamformers in conventional designs are represented as complex-valued matrices with phase-based constraints in the complex domain. This drives the quest for novel optimization strategies for holographic beamforming.

\section{Proposed Solution} \label{section_3}

In this section, we present a novel approach to address the optimization problem \eqref{problem_statement} using an alternating optimization framework. Specifically, we decompose the global problem into three interrelated sub-problems and derive optimal solutions for each variable under the assumption that the other variables remain fixed.

\subsubsection{Digital Beamforming Solution} We assume the scheduled users and the holographic beamformer to be fixed.
Then, the problem of digital beamforming optimization in a multi-user RHS system aims to maximize the signal quality for each user while aiming to eliminate the inter-user interference. Mathematically, the optimization problem for the design the digital beamforming matrix $\mathbf{V}$ can be formulated as 
\begin{subequations} \label{problem_statement_digital}
    \begin{equation}
    \begin{aligned}
          \max_{\mathbf{V}} \quad & \sum_{d=1}^D \log_2 \left( 1 + \frac{\left|x_d \mathbf{h}_d^H \bmW \mathbf{v}_d \right|^2}{\sigma^2 + \sum_{k \neq d} \left| x_k\mathbf{h}_d^H \bmW \mathbf{v}_k \right|^2} \right),
    \end{aligned}
    \end{equation}
    \begin{equation}
        \textbf{s.t.} \quad  \text{Trace}(\bmV \bmV^H) \leq P_{\text{max}}.
    \end{equation}
\end{subequations} 

To solve the optimization problem stated in \eqref{problem_statement_digital}, the challenge lies in the coupled terms within the SINR expressions, as they involve contributions from both the desired signal and inter-user interference. To address this, we adopt the ZF beamforming approach.
It utilizes the pseudo-inverse of the concatenated effective channel matrix $\mathbf{H}_{\text{eff}} = [x_1 \mathbf{h}_1^H \mathbf{W}, x_2 \mathbf{h}_2^H \mathbf{W}, \dots, x_D \mathbf{h}_D^H \mathbf{W}]^H$. Given the effective response including the holographic beamformer, the optimal  ZF beamforming solution is given by

\begin{equation}
\mathbf{V}_{\text{ZF}} = \mathbf{H}_{\text{eff}}^H (\mathbf{H}_{\text{eff}} \mathbf{H}_{\text{eff}}^H)^{-1}.
\end{equation}
Note the the proposed solution considers only the impact of the scheduled users as each effective channel vector is multiplied by the scheduling variables $x_i$.
This ensures that the beamforming vectors are orthogonal to the effective channels of unintended users, thus eliminating inter-user interference. To satisfy the total transmit power constraint, the beamforming matrix is scaled as:

\begin{equation}
\mathbf{V} = \sqrt{\frac{P_{\text{max}}}{\|\mathbf{V}_{\text{ZF}}\|_F^2}} \cdot \mathbf{V}_{\text{ZF}},
\end{equation}

where $\|\mathbf{V}_{\text{ZF}}\|_F^2$ is the Frobenius norm of the unconstrained ZF beamforming matrix. This approach provides a computationally efficient solution while ensuring interference-free communication. However, note that the effective channel depends on $\bmW$, which needs changes at each iteration, and therefore the solution needs to recomputed.

\subsubsection{Holographic Beamforming Optimization}
In the following, we derive the optimal solution for the holographic beamformers, assuming the digital beamforming matrix $\bmV$ and the scheduling binary variables $\bmx$ to be fixed. For such a case, the optimization problem for the holographic beamformer reduces to 

\begin{subequations} \label{problem_statement_holographic}
    \begin{equation}
    \begin{aligned}
          \max_{ \bmW} \quad & \sum_{d=1}^D  \log_2 \left( 1 + \frac{\left|x_d \mathbf{h}_d^H \bmW \mathbf{v}_d \right|^2}{\sigma^2 + \sum_{k \neq d} \left| x_k\mathbf{h}_d^H \bmW \mathbf{v}_k \right|^2} \right),
    \end{aligned}
    \end{equation}
    \begin{equation}
        \textbf{s.t.} \quad  
        0 \leq w_{m} \leq 1,  \quad  \forall\; 0 < m \leq M. 
    \end{equation}
\end{subequations}

Recall that the holographic beamformer has the following structure,  
\{$w_{m} \cdot e^{-j \bmk_s \cdot \bmr_{m}^k}\},$ where $e^{-j \bmk_s \cdot \bmr_{m}^k}$ is the phase of the reference wave from the $k$-th feed. As the phase part is fixed, our objective is to obtain the real-valued weights for each element of the RHS such that the sum rate is maximized. Note that based on the considered holographic architecture, we can decompose the holographic beamformer into two parts as $ \bmW = \text{diag}([w_1, w_2, \dots, w_M]) \mathbf{\Phi} $ \cite{
zhang2022holographic}, where the diagonal elements $ w_m $ represent the adjustable weight of the holographic beamformer for the reconfigurable element $0\leq m \leq M$, and $ \mathbf{\Phi} $ contains fixed phase components, which are fixed according to the reference wave propagation within the waveguide, i.e., with elements of the form $e^{-j \bmk_s \cdot \bmr_{m}^k}$. Each $ w_m $ is constrained to lie in the range $ [0, 1] $, such that it does not amplify the irradiated signal. The optimization focuses on maximizing the sum rate of the RHS-assisted system for the selected users by adjusting these real-valued weights.

To derive the novel optimization design for the holographic beamformer, we first consider highlighting the dependence of the received signal on the weight $w_m$, which allows us to write received signal at user $d \in \mathcal{D}$ denoted as $y_d$ as
\begin{equation}
\begin{aligned}
    y_d = & \sum_{n=1}^M x_d \left(  h_{d,m}^* w_m \sum_{k=1}^K \phi_{m,k} v_{k,d} \right) s_d \\& +\sum_{j \neq d} x_j \sum_{n=1}^M \left( h_{d,m}^* w_m \sum_{k=1}^K \phi_{m,k} v_{k,j} \right) s_j + n_d,
\end{aligned}
\end{equation}
where:
\begin{itemize}
    \item $ h_{d,m} $ is the $ m $-th element of the channel vector for user $ d $,
    \item $ \phi_{m,k} $ is the $ (m,k) $-th element of the fixed phase matrix $ \mathbf{\Phi} $ depending on the $k$-th feed,
    \item $ v_{k,d} $ is the $ k $-th element of the digital beamformer for user $ d $,
    \item $ n_d $ is the noise term.
\end{itemize}

Based on the above decomposition, we can write the SINR as a function of amplitude $w_m$ as 
\begin{equation}
    \text{SINR}_d(w_m) = \frac{  \left| x_d \sum_{m=1}^M w_m h_{d,m}^* \sum_{k=1}^K \phi_{m,k} v_{k,d} \right|^2}{\sigma^2 + \sum_{j \neq d}  \left| x_j \sum_{m=1}^M w_m h_{d,m}^* \sum_{k=1}^K \phi_{m,k} v_{k,j} \right|^2}.
\end{equation}

Given the $\text{SINR}_d(w_m)$, we can restate the optimization problem for the holographic beamformer as a function of the weight $w_m, \forall m$, as 
\begin{subequations}
    \begin{equation}
    \max_{w_m}   \sum_{d \in \mathcal{D}} \log_2 \left( 1 + \text{SINR}_d(w_m) \right),
\end{equation}
\begin{equation}
 \text{s.t} \quad   0 \leq w_m \leq 1, \quad  \forall\; 0 < m \leq M. 
\end{equation}
\end{subequations}

To address this problem, we employ a gradient-ascent approach to directly handle the complex objective function. Specifically, the gradient of the sum rate is calculated with respect to each $w_m$, which is iteratively updated in the direction of the gradient using a gradient-based optimization technique.

Let $R_{\text{sum}}$ denote the total sum-rate. Its gradient with respect to the weight $w_m$ of the reconfigurable element $m$ of RHS is given as
\begin{equation} \label{final_grad}
    \frac{\partial R_{\text{sum}}}{\partial w_m} = \sum_{d \in \mathcal{D}} \frac{1}{\ln(2)} \frac{1}{1 + \text{SINR}_d} \frac{\partial \text{SINR}_d}{\partial w_m}.
\end{equation}

where the last term must be derived. Let $\text{Num}_d$ and  $\text{Den}_d$ denote the numerator and denominator of the  $\text{SINR}_d$, given as

\begin{equation}
   \text{Num}_d=  \left| x_d \sum_{m=1}^M w_m h_{d,m}^* \sum_{k=1}^K \phi_{m,k} v_{k,d} \right|^2,
\end{equation}

\begin{equation}
   \text{Den}_d = \sigma^2 + \sum_{j \neq d}  \left| x_j\sum_{m=1}^M w_m h_{d,m}^* \sum_{k=1}^K \phi_{m,k} v_{k,j} \right|^2.
\end{equation}
To compute $ \frac{\partial \text{SINR}_d}{\partial w_m} $, we use the quotient rule:
\begin{equation}
\begin{aligned}
\frac{\partial \text{SINR}_d}{\partial w_m} =  & \frac{\left(\frac{\partial \text{Num}_d(w_m)}{\partial w_m}\right)  \text{Den}_d(w_m) }{\text{Den}(w_m)^2} \\& - \frac{\text{Num}_d(w_m) \left(\frac{\partial \text{Den}_d(w_m)}{\partial w_m}\right)}{\text{Den}(w_m)^2}.
\end{aligned}
\end{equation}

Differentiating the numerator and the denominator with respect to $ w_m $, we get the following

\begin{equation}
\begin{aligned}
     \frac{\partial \text{Num}_d}{\partial w_m} =  & 2 \text{Re} \left( x_d  \left(\sum_{m=1}^M w_m h_{d,m}^* \sum_{k=1}^K \phi_{m,k} v_{k,d} \right)  \cdot \right. \\ 
     & \left. h_{d,m} \sum_{k=1}^K \phi_{m,k}^* v_{k,d}^* \right).
\end{aligned}
\end{equation}

Differentiating the denominator terms with respect to $ w_m$, leads to the following expression
\begin{equation}
\begin{aligned}
    \frac{\partial \text{Den}_d}{\partial w_m} = & 2 \sum_{j \neq d} \text{Re} \bigg(x_j\left(\sum_{m=1}^M w_m h_{d,m}^* \sum_{k=1}^K \phi_{m,k} v_{k,j}\right) \cdot \\ 
    & h_{d,m} \sum_{k=1}^K \phi_{m,k}^* v_{k,j}^* \bigg).
\end{aligned}
\end{equation}

Based on the aforementioned derivations, the complete gradient of $\frac{\partial \text{SINR}_d}{\partial w_m}$ is given in \eqref{grad}.
Finally, the complete gradient of sum-rate can be obtained by substituting this expression in $\frac{\partial R_{\text{sum}}}{\partial w_m}$ , in equation  \eqref{final_grad}.
\begin{figure*}
   \begin{equation} \label{grad}
   \begin{aligned}
          \frac{\partial \text{SINR}_d}{\partial w_m} = &  \frac{  2 \text{Re} \left(x_d \left(\sum_{m=1}^M w_m h_{d,m}^* \sum_{k=1}^K \phi_{m,k} v_{k,d}\right) \cdot h_{d,m} \sum_{k=1}^K \phi_{m,k}^* v_{k,d}^* \right) \left( \sigma^2 + \sum_{j \neq d}  \left| x_j \sum_{m=1}^M w_m h_{d,n}^* \sum_{k=1}^K \phi_{m,k} v_{k,j} \right|^2 \right)}{\left( \sigma^2 + \sum_{j \neq d}  \left| x_j \sum_{m=1}^M w_m h_{d,m}^* \sum_{k=1}^K \phi_{m,k} v_{k,j} \right|^2 \right)^2} \\& - \frac{ x_d \left| \sum_{m=1}^M w_m h_{d,m}^* \sum_{k=1}^K \phi_{m,k} v_{k,d} \right|^2 2 \sum_{j \neq d} x_j \text{Re}\left( \left(\sum_{m=1}^M w_m h_{d,m}^* \sum_{k=1}^K \phi_{m,k} v_{k,j}\right) \cdot h_{d,m} \sum_{k=1}^K \phi_{m,k}^* v_{k,j}^* \right) }{\left( \sigma^2 + \sum_{j \neq d} x_j \left| \sum_{m=1}^M w_m h_{d,m}^* \sum_{k=1}^K \phi_{m,k} v_{k,j} \right|^2 \right)^2}
   \end{aligned}
\end{equation} \hrulefill
\end{figure*}

To optimize the holographic beamforming weights $ w_m $, we utilize a gradient-ascent-based iterative approach to maximize the RHS network sum-rate. This method ensures that the sum rate is incrementally maximized while adhering to the interval constraints on each element of the RHS $ w_m $. At each iteration, the weights are adjusted by moving in the direction of the gradient of the sum rate with respect to $ w_m $. The gradient provides the steepest ascent direction, ensuring that the sum rate increases with each update.

To enforce the physical constraint that the weight $ w_m $ remains within the range $[0, 1]$, the updated value is projected in the feasible space. The update rule for $ w_m $ at iteration $ t+1 $ is expressed as:

\begin{equation}
    w_m^{(t+1)} = \min \left( 1, \max \left( 0, w_m^{(t)} + \eta \frac{\partial R_{\text{sum}}}{\partial w_m} \right) \right),
\end{equation}

where $ \eta > 0 $ is the learning rate that determines the step size of each update, $ \frac{\partial R_{\text{sum}}}{\partial w_m} $ is the partial derivative of the sum rate $ R_{\text{sum}} $ with respect to $ w_m $, calculated based on the system's SINR expressions. The $\max(0, \cdot)$ operation ensures that the updated weights do not fall below zero. The $\min(1, \cdot)$ operation ensures that the updated weights do not exceed one.

For a fixed digital beamforming matrix $ \bmV $ and scheduling decision vector $ \bmx $, the iterative procedure is repeated until convergence is achieved. Convergence is defined as the point at which the change in the sum rate between consecutive iterations falls below a predefined threshold, or when the maximum allowable number of iterations is reached. The detailed steps of the complete algorithm are formally presented in Algorithm 1.

\begin{algorithm}[t]
\caption{Gradient-Based Optimization for Holographic Beamforming}
\begin{algorithmic}[1] \label{Holographic_alg}
\State \textbf{Input:} Channel vectors $ \mathbf{h}_d $, phase matrix $ \mathbf{\Phi} $, digital beamforming matrix $ \mathbf{V} $, user scheduling vector $ \mathbf{x} $, noise variance $ \sigma^2 $, and maximum iterations $ T_{\text{max}} $.
\State \textbf{Initialize} $ w_m^{(0)} $ for all $ m $, learning rate $ \eta > 0 $, and convergence threshold $ \epsilon $.
        \State Compute the sum rate:
    \[
    R_{\text{sum}}^{(t)} \gets \sum_{d \in \mathcal{D}} \log_2 \left( 1 + \text{SINR}_d(w_m^{(t)}) \right)
    \]
     \State \textbf{for m = 1...M}
    \State  Compute and update $ w_m $ using gradient ascent:
        \[
        w_m^{(t+1)} \gets \min \left( 1, \max \left( 0, w_m^{(t)} + \eta \frac{\partial R_{\text{sum}}}{\partial w_m} \right) \right)
        \] \newline
        \textbf{Check Convergence with updated rate}   \newline
     \textbf{If} $ \big| R_{\text{sum}}^{(t+1)} - R_{\text{sum}}^{(t)} \big| < \epsilon $, \textbf{then} Stop \newline
    \State \hspace{0.1cm}\textbf{else}  Go to step $3$ \newline
     \State \textbf{end}
\State \textbf{Return:} Optimized weights $ \{w_m\} $. 
\end{algorithmic}
\end{algorithm}

 \subsubsection{User Scheduling for an RHS Assisted System}
In this section, we address the problem of optimizing user scheduling, given the fixed digital beamforming matrix $\mathbf{V}$ and holographic beamforming matrix $\mathbf{W}$ from the previous iteration, the optimization problem can be formally expressed as  
 
\begin{subequations} \label{problem_statement_user_selection}
    \begin{equation}
    \begin{aligned}
          \max_{\mathbf{x}} \quad & \sum_{d=1}^D   \log_2 \left( 1 + \frac{\left|x_d \mathbf{h}_d^H \bmW \mathbf{v}_d \right|^2}{\sigma^2 + \sum_{k \neq d} \left| x_k\mathbf{h}_d^H \bmW \mathbf{v}_k \right|^2}  \right),
    \end{aligned}
    \end{equation}
    \begin{equation}
        \textbf{s.t.} \quad x_d \in \mathcal{S}, \quad \forall d \in \mathcal{D}
    \end{equation}
    \begin{equation}
       \log_2 \left( 1 + \frac{\left|x_d \mathbf{h}_d^H \bmW \mathbf{v}_d \right|^2}{\sigma^2 + \sum_{k \neq d} \left| x_k\mathbf{h}_d^H \bmW \mathbf{v}_k \right|^2}  \right) \geq R_d^{\text{min}}, \; \forall x_d =1.
    \end{equation}
\end{subequations} 

This problem, however, is inherently complex due to the combinatorial nature of user selection and the interdependence of users' QoS constraints and interference dynamics.  The optimization problem aims to maximize the sum rate by determining the optimal scheduling vector $\mathbf{x}$, where each element $x_d$ is a binary variable indicating whether user $d$ is scheduled ($x_d = 1$) or not ($x_d = 0$). This decision must satisfy the constraints that each selected user achieves their required minimum rate, $R_d^{\text{min}}$, while accounting for the interference generated by other users. To efficiently solve this problem, we propose a greedy algorithm that iteratively selects users based on their channel conditions and contribution to the overall system performance, including also interference generation towards other users. This approach balances computational efficiency with optimality, making it well-suited for large-scale RHS-assisted networks.

To design the greedy user scheduling algorithm, we first assume that all users are active ($x_d = 1$, $\forall d \in \mathcal{D}$). Based on the initialization of the digital beamformers and the holographic beamformer, the achievable rates for each user is calculated, which can be expressed as  
\begin{equation}
R_d = \log_2 \left( 1 + \frac{\left| \mathbf{h}_d^H \mathbf{W} \mathbf{v}_d \right|^2}{\sigma^2 + \sum_{k \neq d} \left| \mathbf{h}_d^H \mathbf{W} \mathbf{v}_k \right|^2} \right).
\end{equation}

By following the greedy principle, the users are identified in terms of achieved sum-rate, which are then sorted in descending order based on their rates, prioritizing those with better channel conditions and greater potential to contribute to maximizing the RHS-assisted system's sum rate. Note that as the digital beamformers and the holographic beamformer are not optimized, the effective per-user rate is different than the initial guess of the per-user rate. Therefore, we cannot exclude certain users solely based on whether they meet the QoS constraint based on the initial guess.

\begin{algorithm}[t]
\caption{Joint Holographic Beamforming and Greedy User Scheduling}
\begin{algorithmic}[1]
\State \textbf{Input:} Given the channel responses, initialize the holographic beamforming matrix $\mathbf{W}$, digital beamforming matrix $\mathbf{V}$, set the minimum QoS constraint $\mathbf{R}^{\text{min}}$.
\State Compute initial rates for all users assuming all are active:
\[
R_d = \log_2 \left( 1 + \frac{\left| \mathbf{h}_d^H \mathbf{W} \mathbf{v}_d \right|^2}{\sigma^2 + \sum_{k \neq d} \left| \mathbf{h}_d^H \mathbf{W} \mathbf{v}_k \right|^2} \right).
\]
\State Sort eligible users by $R_d$ in descending order.
    \State \textbf{for} d in the sorted list
    \State \hspace{2mm} Tentatively select user $d$ by setting $x_d = 1$.
    \State \hspace{2mm} Update $\bmW$ with Algorithm $1$.
    \State \hspace{2mm}  Update the digital beamformer $\mathbf{V}$ using ZF.
    \State \hspace{2mm} Recalculate rates for all selected users:
    \[
    R_d = \log_2 \left( 1 + \frac{\left| \mathbf{h}_d^H \mathbf{W} \mathbf{v}_d \right|^2}{\sigma^2 + \sum_{k \neq d, x_k=1} \left| \mathbf{h}_d^H \mathbf{W} \mathbf{v}_k \right|^2} \right).
    \]
\hspace{2mm}   \textbf{If} $R_d \geq R_d^{\text{min}}, \, \forall d \text{ such that } x_d = 1$
        \State \hspace{3mm}Accept user $d$: Update $x_d = 1$.\\
 \hspace{2mm}   \textbf{Else}  Reject user $d$: Update $x_d = 0$..
    \EndIf
\State \textbf{end}
\State \textbf{Return:} $\mathbf{x}$, $\bmV$ and $\bmW$.
\EndIf
\end{algorithmic}
\end{algorithm}

Once the users are sorted in descending order, we proceed with an iterative process to determine the optimal scheduling vector, by jointly also updating the digital and the holographic beamformers to account for the effective interference. For the sorted list, the algorithm tentatively selects the first user by setting its scheduling variable $x_d =1$. As there is no interference, the digital beamformers and the holographic beamformer are updated to maximize the received power. Subsequently, user $d$ is tentatively added to the set of feasible users. Following this tentative selection, the holographic beamforming matrix $\mathbf{W}$ is updated using a gradient ascent method to refine the holographic patterns. This step ensures that the interference is minimized, and power is effectively directed toward the selected users. Concurrently, the digital beamforming matrix $\mathbf{V}$ is updated using the ZF approach to further suppress the inter-user interference and optimize the signal quality for each selected user. After updating the beamformers, we recalculate the achievable rates for all currently scheduled users to verify that the QoS constraints are met. If the recalculated rates confirm that all selected users satisfy their minimum rate requirements, the tentative inclusion of user $d$ is finalized, and the scheduling vector is updated accordingly. If any QoS constraint is violated, the algorithm sets the scheduling decision variable for that user $x_d =0$ and moves to the next users in the sorted list. This iterative process continues until all eligible users have been evaluated which satisfies the minimum QoS while providing their maximum contribution to maximize the network sum-rate. The complete algorithm which performs joint hybrid holographic beamforming and user scheduling under the QoS constraint is formally stated in Algorithm $2$.

It is noteworthy that although ZF beamforming leads to zero-force inter-user interference, the solution is calculated based on the effective channel response, which depends on the holographic beamformer, and the scheduled users. Therefore, at each tentative update of the scheduling vector, the digital beamformers and the holographic beamformer need to be updated to evaluate the effective impact of the new user in the set of feasible users. This joint design leads to significant performance gains in terms of sum-rate maximization as the algorithm follows the greedy principle and prioritizes the users with the potential to contribute the most to the global RHS-assisted network sum-rate improvement.

\subsection{On the Convergence}
The convergence of the proposed alternating optimization algorithm for maximizing the sum rate in a multi-user system is established based on fundamental principles of optimization theory. The algorithm addresses the non-convex optimization problem by decomposing it into three subproblems: digital beamforming, holographic beamforming, and user selection, which are solved iteratively in a cyclic manner. At each iteration, the digital beamforming subproblem is solved using the ZF technique, which provides an optimal solution for the fixed holographic beamformer and user set. The holographic beamforming weights are updated using a gradient ascent method constrained within the compact feasible set $[0, 1]$, ensuring that the update step is well-defined and feasible. The user selection subproblem employs a greedy optimization approach that incrementally improves the objective function while satisfying the minimum rate constraints. For each user scheduling update, the digital beamformers and holographic beamformer are iteratively updates thus ensuring that the sum rate is maximized.

The convergence of the algorithm is guaranteed under the following conditions. First, at each iteration, the sum rate $ R_{\text{sum}} $ is monotonically non-decreasing, as each subproblem is solved to either improve or maintain the objective value. Second, the sum rate $ R_{\text{sum}} $ is bounded above due to practical constraints such as limited power and finite channel gains. Therefore, the sequence of objective values generated by the algorithm is monotonically increasing and bounded, implying convergence to a finite value. Additionally, the alternating optimization framework ensures that, at convergence, the gradients of $ R_{\text{sum}} $ with respect to all variables vanish or satisfy the Karush-Kuhn-Tucker (KKT) conditions for the respective subproblems. Consequently, the algorithm converges to a stationary point, which represents a locally optimal solution to the original non-convex problem.

\subsection{Computational Analysis}

The joint algorithm combines digital beamforming, holographic beamforming, and user selection, with iterative updates for each selected user. Assuming $D$ is the total number of users and $D'$ is the number of users ultimately selected ($D' \leq D$), the computational cost for each component is derived below.

The digital beamforming step uses the ZF approach to cancel inter-user interference. For a given set of $D'$ selected users, the effective channel matrix $\mathbf{H}_{\text{eff}} \in \mathbb{C}^{D' \times M}$ must be constructed, which requires $O(D' \cdot M)$ operations. The computation of $\mathbf{H}_{\text{eff}} \mathbf{H}_{\text{eff}}^H \in \mathbb{C}^{D' \times D'}$ has complexity $O(D'^2 \cdot M)$, and inverting this $D' \times D'$ matrix incurs $O(D'^3)$. The final multiplication to compute the ZF beamforming matrix has complexity $O(M \cdot D'^2)$, while scaling the matrix adds $O(M \cdot D')$. The dominant term in a single digital beamforming update is $O(\max(D'^3, D'^2 \cdot M))$, and this step must be repeated for each selected user during the iterative user selection process.

The holographic beamforming step employs gradient-based optimization to update the beamforming weights for $M$ holographic elements. For $D'$ users, the SINR gradient computation involves $O(M \cdot K \cdot D')$ operations for a single gradient. Since there are $M$ weights, the total cost per iteration is $O(M^2 \cdot K \cdot D')$. With $T_{\text{max}}$ iterations of gradient descent, the overall complexity of holographic beamforming for a single update is $O(T_{\text{max}} \cdot M^2 \cdot K \cdot D')$. Like digital beamforming, this step is also performed for each selected user.

The user selection process starts by computing the initial SINR for all $D$ users, which involves $O(D \cdot M \cdot K)$ operations. Sorting the eligible users incurs $O(D \log D)$. During the iterative selection process, $D'$ users are chosen, with each iteration requiring updates to both the digital and holographic beamformers, followed by recalculations of the SINR for the selected users.

Combining these components, the total complexity of the joint algorithm is given by:
\[
O(D \cdot M \cdot K + D \log D + D' \cdot (T_{\text{max}} \cdot M^2 \cdot K \cdot D' + \max(D'^3, D'^2 \cdot M))).
\]
 
 The term $D' \cdot T_{\text{max}} \cdot M^2 \cdot K \cdot D'$ represents the computational cost of the iterative updates to the holographic beamformer for each selected user, scaling with the number of iterations, holographic elements, and selected users. Similarly, the term $D' \cdot \max(D'^3, D'^2 \cdot M)$ captures the cost of repeatedly updating the digital beamformer, which is driven by the complexity of constructing and inverting the effective channel matrix for the selected users. In comparison, the initial SINR calculations and sorting, with a combined complexity of $O(D \cdot M \cdot K + D \log D)$, are relatively minor contributors, particularly for large numbers of selected users ($D'$), holographic beamforming elements ($M$), or iterations ($T_{\text{max}}$).

For large-scale systems where $D' \approx D$, the overall complexity simplifies to:
\[
O(D \cdot (T_{\text{max}} \cdot M^2 \cdot K + \max(D^3, D^2 \cdot M))).
\]

This expression highlights the computational cost of the iterative updates for both digital and holographic beamformers, scaled by the number of users ultimately selected $D'$, the beamforming dimensions $M$, and the gradient descent iterations $T_{\text{max}}$.

\section{Simulation Results} \label{section_4}
In this section, we present simulation results to evaluate the performance of the proposed algorithms for joint user scheduling and hybrid holographic beamforming.

We assume that the RHS-assisted QoS-constrained system has a total of $ D = 6 $ users, and the base station has $R=8$ RF chains. The RHS is assumed to be either of size $6 \times 6= 36$ or $8 \times 8= 64$ for holographic weight-controlled beamforming. The total transmit power at the base station is set to $ P_{\text{max}} = 1 W$, and a minimum rate requirement for the scheduled users is set to be $ R_d^{\text{min}}=R_{\text{min}} = 5 bps/Hz $. We define the SNR of the system as the transmit SNR, i.e., $ \text{SNR} = P_{max}/\sigma_b,$ with $\sigma_b$ 
denoting the noise variance at the base station. The base station and the users are assumed to have the same level of noise variance, i.e. $\sigma_b = \sigma_d, \forall d$, which is set to meet the desirable SNR values given the total transmit power. The holographic beamformer employs iterative amplitude optimization, for which the convergence threshold of $ \epsilon = 10^{-5} $, and a learning rate of $ \eta = 0.01 $ are chosen.  The carrier frequency of $30$GHz and the element spacing $\lambda/3$ is assumed. Recall that the propagation vector in free space is represented by $ \mathbf{k}_f $, while the propagation vector on the RHS is denoted as $ \mathbf{k}_s $. According to electromagnetism, the magnitude of $ \mathbf{k}_s $ is related to $ \mathbf{k}_f $ by the equation
$|\mathbf{k}_s| = \sqrt{\epsilon_r} |\mathbf{k}_f|$,
where $ \epsilon_r $ is the relative permittivity of the RHS substrate, typically around $3$. We set these parameters with the typical values used in the literature as $|\mathbf{k}_s| = 200 \sqrt{3} \pi $ and $|\mathbf{k}_f| = 200  \pi$ \cite{deng2022reconfigurable}. 

\begin{figure*}
    \centering
 \begin{minipage}{0.49\textwidth}
      \centering
    \includegraphics[width=0.9\linewidth]{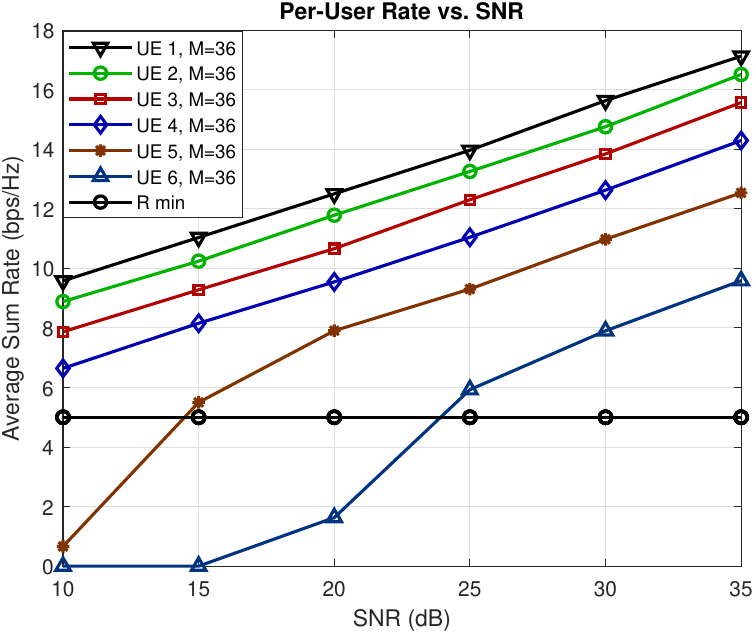}
    \caption{Per user Sum-rate for RHS system with QoS constraint as a function of SNR.}
     \label{fig2}
\end{minipage}  
      \begin{minipage}{0.49\textwidth}
      \centering
    \includegraphics[width=0.9\linewidth]{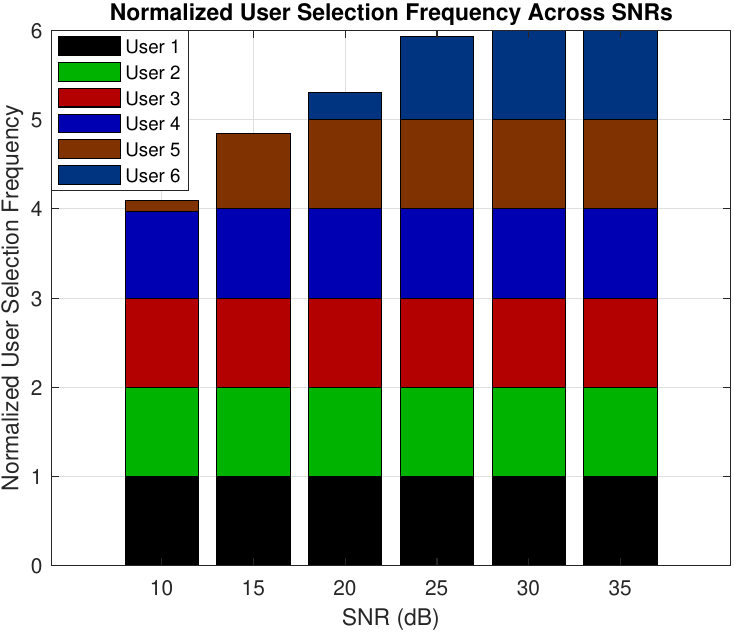}
    \caption{Normalized user selection frequency with M$=36$.}
    \label{fig3}
    \end{minipage}  
\end{figure*}
\begin{figure*}
    \centering
 \begin{minipage}{0.49\textwidth}
      \centering
    \includegraphics[width=0.9\linewidth]{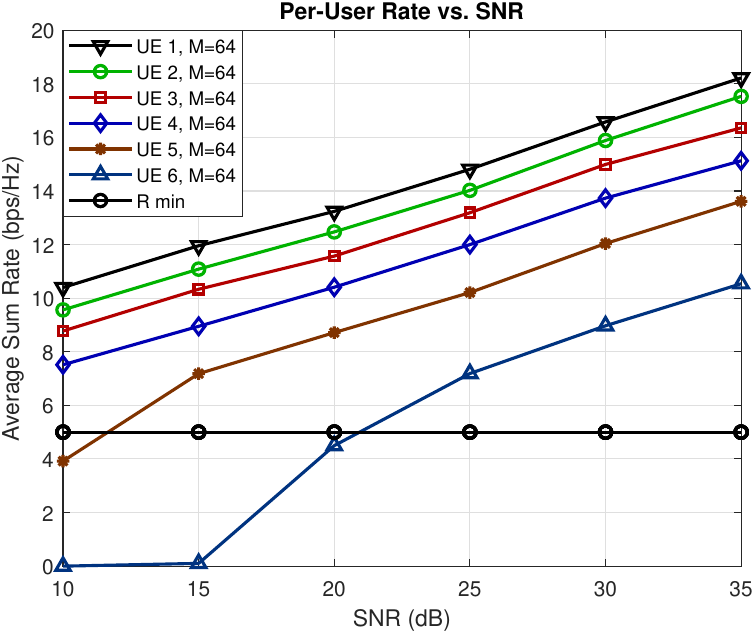}
    \caption{Per user rate for an RHS-assisted system with QoS constraint as a function of SNR with M=$64$.}
    \label{fig4}
\end{minipage}  
      \begin{minipage}{0.49\textwidth}
      \centering
    \centering
    \includegraphics[width=0.9\linewidth]{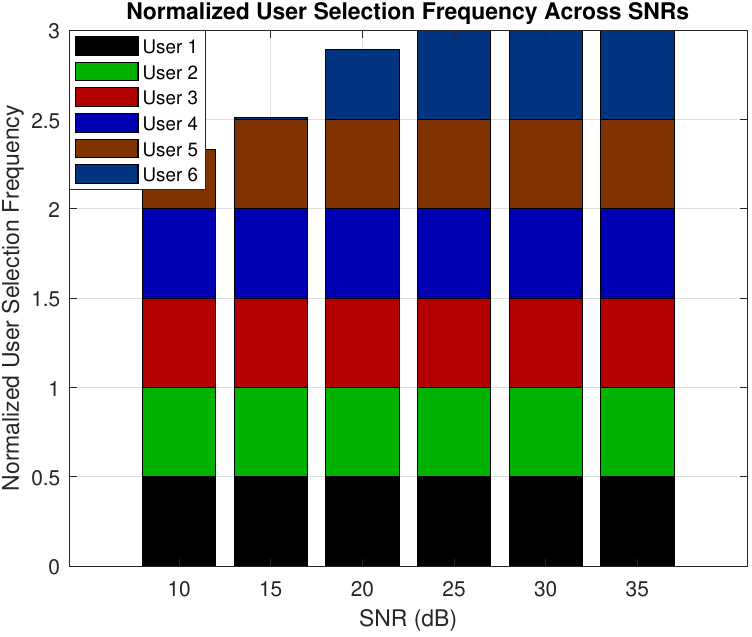}
    \caption{Normalized user selection frequency with M$=64$.}
    \label{fig5}
    \end{minipage}  
\end{figure*}

\begin{figure*}
    \centering
 \begin{minipage}{0.49\textwidth}
      \centering
    \includegraphics[width=0.9\linewidth]{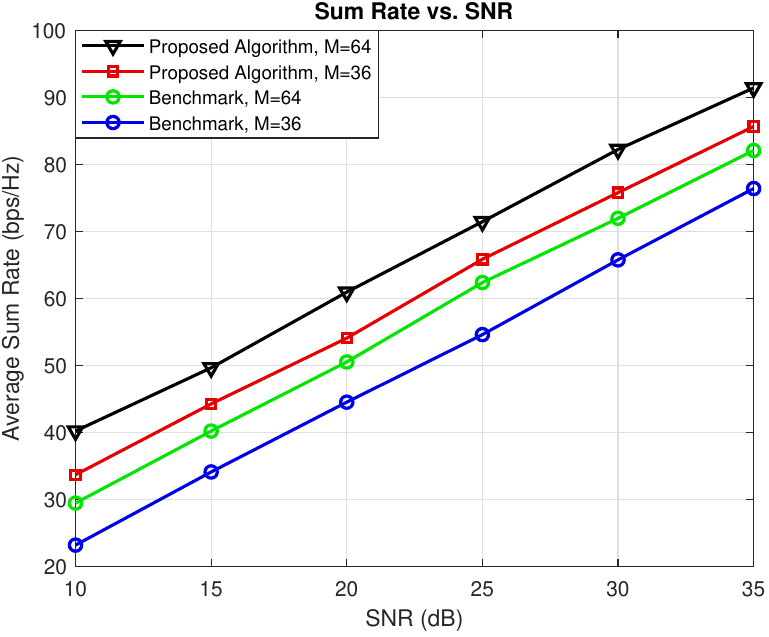}
    \caption{Average Sum-rate as a function of SNR with different with M$=36$ and M$=64$ in comparison with the benchmark scheme.}
    \label{fig6}
\end{minipage}  
      \begin{minipage}{0.49\textwidth}
      \centering
    \includegraphics[width=0.9\linewidth]{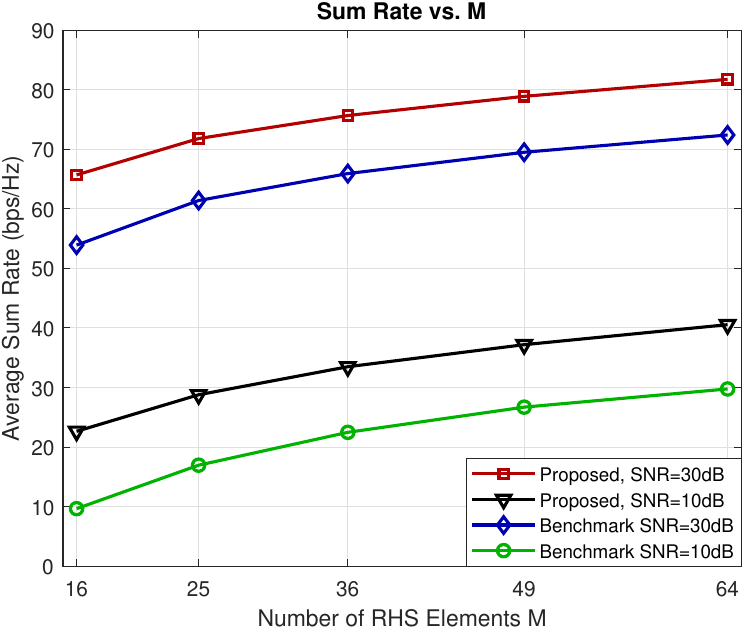}
    \caption{Average Sum-rate as a function of RHS size with SNR$=30$dB and SNR$=10$dB in comparison with the benchmark scheme.}
    \label{fig7}
    \end{minipage}  
\end{figure*} 
Additionally, for the 6 users in the system, we introduce variations in their channel responses by scaling them with a factor, ensuring that some users are more favourable for maximizing the network sum rate than others. The users are arranged in ascending order, with user $i$ having a channel gain 20\% greater than that of user $i+1$. Consequently, the scale factor vector is defined as
$\bm{f} = [1.2, 1, 0.8, 0.6, 0.4, 0.2]^T,$ where the scale factor in position $i$ corresponds to user $i$. Note that users 5 and 6 experience the worst channel conditions compared to the first four users. To assess the performance of the proposed design, the first part of the simulations defines the benchmark as the minimum QoS requirement for each user, denoted as $R_{\text{min}}$, and analyzes the per-user performance in terms of achievable rate and scheduling frequency given the diverse channel conditions. In the second part, the benchmark scheme optimizes the digital beamformers and scheduling variables, but the amplitudes of the holographic beamformers are randomly selected within the interval $[0, 1]$. For the simulations, the results are averaged over $100$ channel realizations.

Figure \ref{fig2} shows the average per-user rate versus SNR for the $6$ users with RHS consisting of $36$ elements, with each user experiencing different channel conditions. As expected, users with higher channel gains, such as User 1, perform significantly better compared to users with poorer channel conditions, such as User $5$ and Users $6$. Note that we are plotting the average rate over multiple channel realizations. Therefore, users whose rates exceed the average minimum QoS requirement $ R_{\text{min}} = 5 \text{ bps/Hz} $ are scheduled more often. However, at low SNR, the number of times that Users $5$ and $6$ are not scheduled for transmission increases, which results in their average rates being lower than $ R_{\text{min}} $. In contrast, users with better channel conditions, such as Users $1$-$4$, are more frequently scheduled and are thus able to achieve an average rate that meets or exceeds the minimum QoS requirement. The plot clearly shows that as the SNR increases, the per-user rate improves for all users. For instance, at an SNR of $35$ dB, User 1 achieves the highest rate, nearing $17$ bps/Hz, while User $6$ continues to experience a lower average rate. This trend is consistent with the assumption that users with better channel conditions will achieve higher rates, as the proposed approach is greedy and aims to maximize the network sum-rate without concern about fairness among user scheduling. Therefore, users with poorer channel conditions, such as User $5$ and User $6$, struggle to achieve high rates even as the SNR increases. For further validation, in Figure \ref{fig3} we plot the normalized user scheduling frequency across different SNRs. The figure reveals that, at lower SNRs, users with better channel conditions, such as Users $1$–$4$, are scheduled more frequently than those with poorer conditions, such as Users $5$ and $6$. This scheduling pattern aligns with the approach that maximizes the network sum-rate, prioritizing users with higher channel gains. As the SNR increases, the scheduling frequency for all users becomes more uniform at high SNR, reflecting the reduced impact of channel variations as the system improves its overall performance.

Figures \ref{fig4} and \ref{fig5} present the same results but with the RHS size increased to 64. As the RHS size increases, there is a noticeable improvement in the average per-user rate. Specifically, for User 6 at an SNR of 20 dB, it is evident that despite experiencing the worst channel conditions, a larger RHS size leads to better performance and an increased number of scheduling instances, as compared to the results shown in Figure \ref{fig2}. Additionally, Figure \ref{fig5} demonstrates that at low SNR levels, the normalized user scheduling frequency improves for all users, highlighting the beneficial effect of a larger RHS size in enhancing system performance with QoS constraints.

Figure \ref{fig6} illustrates the average sum rate versus SNR for the proposed algorithm and benchmark schemes, with RHS sizes of $64$ and $36$. The results show that the proposed algorithm consistently outperforms the benchmark scheme across all SNR values, with the network sum rate increasing as the SNR improves. Specifically, when the RHS size is increased from $36$ to $64$, the average sum rate increases, indicating the benefit of a larger RHS in improving system performance. This trend is visible for both the proposed algorithm and the benchmark scheme. At lower SNR values (e.g., $10$-$15$ dB), the difference in performance between the proposed algorithm and the benchmark scheme is more pronounced, with the proposed algorithm achieving higher sum rates, especially when the RHS size is $64$. As the SNR increases, the gap in performance between the proposed and benchmark schemes narrows slightly, but the proposed algorithm still maintains a higher sum rate.

Figure \ref{fig7} presents the average sum rate as a function of the number of RHS elements, $ M $, at two distinct SNR levels: $30$ dB and $10$ dB, comparing the proposed algorithm with the benchmark scheme. At an SNR of $30$ dB, the proposed algorithm consistently achieves a higher average sum rate compared to the benchmark scheme, and this gap increases as the number of RHS elements increases. This trend is expected, as a larger RHS allows for more flexibility in holographic beamforming, improving the overall network performance. For the proposed algorithm, the sum rate increases smoothly as $ M $ grows, indicating that more RHS elements contribute to better system performance. At an SNR of $10$ dB, the performance of both the proposed algorithm and the benchmark scheme is lower compared to the $30$ dB scenario, but the overall trend remains the same. The proposed algorithm still outperforms the benchmark and the gap between the two increases with larger $ M $. However, at this lower SNR, the sum rate improvements are more gradual, highlighting the diminished effect of additional RHS elements at lower SNR values. Despite this, the proposed algorithm shows better scalability with the increase in RHS size compared to the benchmark.
\begin{figure*}
    \centering
 \begin{minipage}{0.49\textwidth}
      \centering
    \includegraphics[width=\linewidth]{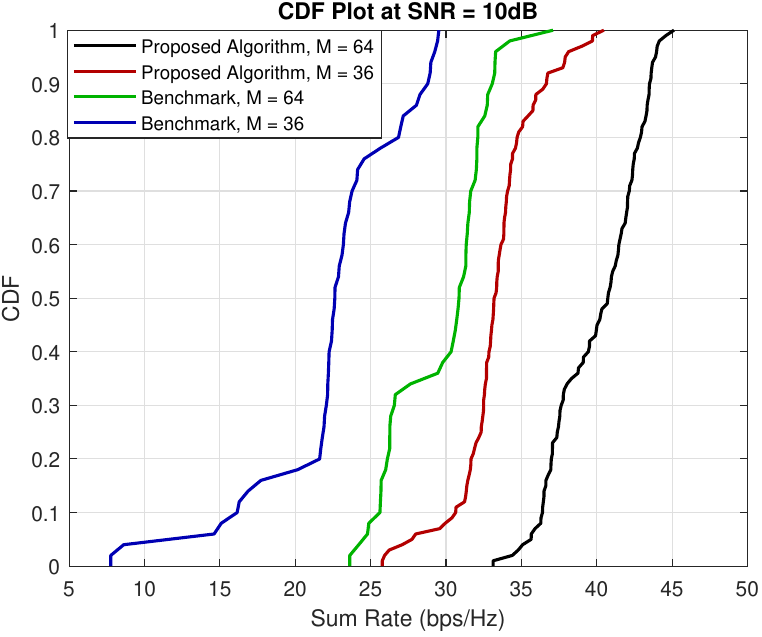}
    \caption{CDF of the proposed algorithm in comparison with the benchmark scheme at SNR$=10$ dB.}
    \label{fig8}
\end{minipage}  
      \begin{minipage}{0.49\textwidth}
      \centering
    \includegraphics[width=\linewidth]{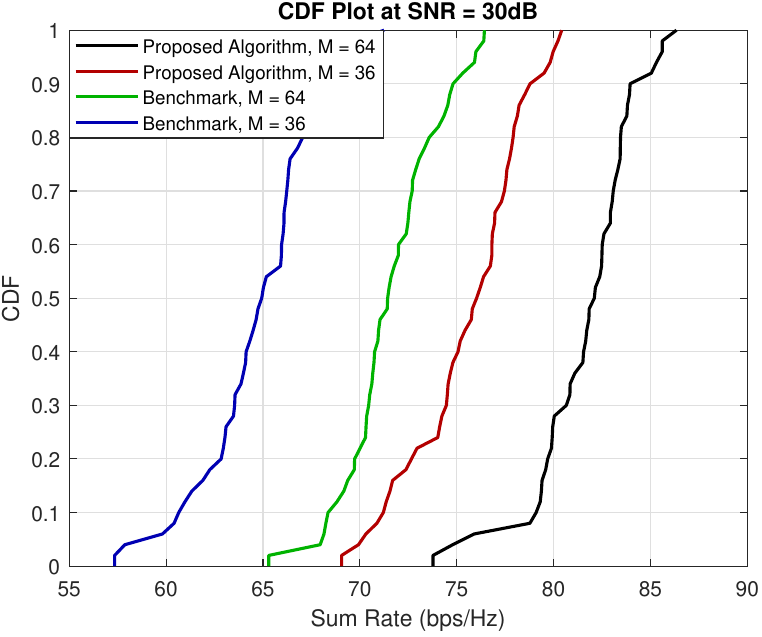}
    \caption{CDF of the proposed algorithm in comparison with the benchmark scheme at SNR$=30$ dB.}
    \label{fig9}
    \end{minipage}  
\end{figure*}

Figure \ref{fig8} presents the cumulative distribution function (CDF) at an SNR of 10 dB for the proposed algorithm and the benchmark scheme with different RHS sizes. The CDF represents the probability that the system’s sum rate will be less than or equal to a given value. Essentially, it shows the fraction of realizations that achieve a particular sum rate or lower. From this plot, it is evident that the proposed algorithm with an RHS size of 64 outperforms the benchmark scheme with the same RHS size across the entire range of sum rates. Specifically, the proposed algorithm reaches a CDF of $1$ (i.e., $100\%$ of the realizations) at a sum rate of around $45$ bps/Hz, whereas the benchmark scheme achieves the same CDF at a significantly lower sum rate of approximately $35$ bps/Hz for an RHS of size $64$. 
For the proposed algorithm with RHS size $36$, although the sum rate is lower than that achieved with RHS size $64$, it still outperforms the benchmark scheme with the same number of RHS elements. The proposed algorithm with RHS size 36 achieves a CDF of $1$ at a sum rate near $40$ bps/Hz, while the benchmark with the same RHS size reaches the same CDF at only around $28$ bps/Hz. This indicates that, even with fewer RHS elements, the proposed algorithm performs better than the benchmark. Figure \ref{fig9} shows the CDF at a higher SNR of $30$ dB for both the proposed algorithm and the benchmark scheme with different RHS sizes. At this higher SNR, the overall sum rates are considerably higher than in the 10 dB scenario, reflecting the increased signal quality. The proposed algorithm with RHS size $64$ continues to outperform the benchmark scheme, reaching a CDF of $1$ at a sum rate of approximately $87$ bps/Hz, compared to the benchmark's $77$ bps/Hz. Similarly, the proposed algorithm with RHS size $36$ achieves a CDF of $1$ at around $81$ bps/Hz, whereas the benchmark reaches the same CDF at approximately $70$ bps/Hz. This demonstrates that the proposed algorithm exhibits superior scalability with RHS size, especially at higher SNR values. In summary, the CDF plots clearly show that both the increase in RHS size and the higher SNR contribute to substantial improvements in system performance. The proposed algorithm consistently provides higher sum rates compared to the benchmark scheme across different RHS sizes and SNR levels, with larger RHS sizes and higher SNRs further enhancing its performance.

The simulation results clearly demonstrate the effectiveness of the proposed algorithms for joint user scheduling and hybrid holographic beamforming in a RHS-assisted QoS-constrained system. As the RHS size increases from $ M = 36 $ to $ M = 64 $, there is a noticeable improvement in system performance, both in terms of the average per-user rate and network sum rate, particularly at higher SNR values. The proposed algorithm consistently outperforms the benchmark scheme across all test cases, including various RHS sizes and SNR levels. The advantage of the proposed approach is evident from the improved scheduling frequencies for users with better channel conditions, which results in higher achievable rates, particularly for users with favorable channel conditions such as Users $ 1 $–$ 4 $.

The CDF analysis at both $ \text{SNR} = 10 \, \text{dB} $ and $ \text{SNR} = 30 \, \text{dB} $ further validates these findings. The CDF plots show that the proposed algorithm, with larger RHS sizes, consistently achieves higher sum rates across all realizations compared to the benchmark. Specifically, at $ \text{SNR} = 30 \, \text{dB} $, the proposed algorithm with RHS size $ M = 64 $ achieves a sum rate of around $ 80 \, \text{bps/Hz} $, outperforming the benchmark by a substantial margin. At lower $ \text{SNR} $ ($ \text{SNR} = 10 \, \text{dB} $), the performance gap remains, though it is smaller due to the overall reduction in signal quality. The results highlight the scalability of the proposed algorithm with increasing RHS size, which enhances system performance by improving beamforming flexibility and ensuring better scheduling decisions.

In summary, the simulations demonstrate that the proposed algorithm not only meets the minimum QoS requirements more effectively but also maximizes the network sum rate by making optimal use of the RHS elements. Moreover, the results suggest that both increasing the RHS size and improving the $ \text{SNR} $ are crucial factors for achieving superior performance in systems relying on holographic beamforming.

\section{Conclusions} \label{section_5}
This work explored the challenges and a solution associated with joint holographic beamforming optimization and user scheduling under QoS constraints in an RHS-assisted system. The problem of maximizing the system's sum-rate while ensuring each user meets its minimum QoS requirement was formulated as a mixed-integer non-convex optimization problem. To address this challenge, a novel algorithmic framework that decomposes the optimization task into three key sub-problems is proposed, which dynamically selects users based on their channel conditions and their potential to maximize the network sum-rate, while ensuring that each user’s QoS constraint is satisfied. Through extensive simulations, the effectiveness of the proposed algorithm is validated at different SNR levels and with different RHS sizes. The results suggest that the joint optimization of holographic beamforming and user scheduling can unlock the full potential of RHS-assisted systems with limited resources and stringent QoS requirements, providing a scalable and efficient solution for next-generation RHS-assisted wireless communication networks.

\ifCLASSOPTIONcaptionsoff
  \newpage
\fi

{\footnotesize
\bibliographystyle{IEEEtran}
\bibliography{main}}
  
\end{document}